\documentstyle[aps,prl]{revtex}
\begin{document}

\title{Soliton structure dynamics in inhomogeneous media}
\author{L. E. Guerrero\thanks{Corresponding author. Fax: +582-9063601; e-mail: lguerre@usb.ve}}
\address{Departamento de F\'{\i}sica, Universidad Sim\'on Bol\'{\i}var, Apartado Postal 89000,
Caracas 1080-A, Venezuela}
\author{A. Bellor\'{\i}n}
\address{Departamento de F\'{\i}sica, Universidad Sim\'on Bol\'{\i}var, Apartado Postal 89000, Carac
as 1080-A, Venezuela}
\author{J. A. Gonz\'alez}
\address{Centro de F\'{\i}sica, Instituto Venezolano de Investigaciones
Cient\'{\i}ficas, Apartado Postal 21827, Caracas 1020-A, Venezuela}
\date{\today}
\maketitle
\begin{abstract}
We show that soliton interaction with finite-width inhomogeneities can activate a 
great number of soliton internal modes. We obtain the exact stationary soliton solution
in the presence of inhomogeneities and solve exactly the stability problem. We present 
a Karhunen-Lo\`eve analysis of 
the soliton structure dynamics as a time-dependent force pumps energy into the
traslational mode of the kink. We
show the importance of the internal modes of the soliton as they can generate 
shape chaos for the soliton as well as cases in which the first shape mode leads the 
dynamics.
\end{abstract}
\pacs{02.30.Jr; 05.45.+b; 52.35.Mw; 52.35.Sb}

\section{Introduction}

The propagation of solitons in the presence of inhomogeneities concerns a
wide variety of condensed matter systems. The traditional approach considers
structureless solitons and delta-function-like impurities.

Real scenarios involve finite-width impurities and under certain
circunstances, the extended character of the soliton must be considered\cite
{r:1,r:2,t:1,r:3}. For instance, the lenght scale competition between the width
of inhomogeneities, the distance between them and the width of the
kink-soliton leads to interesting phenomena like soliton explosions\cite{r:2}%
.

In this paper we take into account the extended character of both the
soliton and the impurity and show that these considerations lead to the
existence of a finite number of soliton internal modes that underlies a rich
spatiotemporal dynamics. We present a model for which the exact stationary
soliton solution in the presence of inhomogeneities can be obtained and the
stability problem can be solved exactly. We use the Karhunen-Lo\`eve (KL)
decomposition to relate the excitation of soliton internal modes with the
sequence of bifurcations obtained as the amplitude of a space-time-dependent
driving force (fitted to the shape of the translational mode) is increased.

\section{The model}

The topological solitons studied in the present paper possess important
applications in condensed matter physics. For instance, in solid state
physics, they describe domain walls in ferromagnets or ferroelectric
materials, dislocations in crystals, charge-density waves, interphase
boundaries in metal alloys, fluxons in long Josephson junctions and
Josephson transmission lines, etc.\cite{s:1,s:2}

Although some of the above mentioned systems are described by the $\phi ^4$%
-model and others by the sine-Gordon equation (and these equations, in their
unperturbed versions, present differences like the fact that the sine-Gordon
equation is completely integrable whereas the $\phi ^4$-model is not) the
properties of the solitons supported by sine-Gordon and $\phi ^4$ equations
are very similar. In fact, these equations are {\it topologically equivalent}
and very often the result obtained for one of them can be applied to the
other\cite{s:1}.

Here we consider the $\phi ^4$ equation in the presence of inhomogeneities
and damping:

\begin{equation}
\phi _{xx}-\phi _{tt}-\gamma \phi _t+\frac 12\left( \phi -\phi ^3\right)
=-N(x)\phi -F(x),  \label{a:1}
\end{equation}
where $F(x)$ is a function with (at least) one zero and $N(x)$ is a
bell-shaped function that rapidly decays to zero for $x\rightarrow \pm
\infty $. An impurity of the kind $N(x)\phi $, but using delta functions,
has been presented in Ref. \cite{r:4}.

In ferroelectric materials $\phi $ is the displacement of the ions from
their equilibrium position in the lattice, $\frac 12\left( \phi -\phi
^3\right) $ is the force due to the anharmonic crystalline potential, $F(x)$
is an applied electric field, and $N(x)$ describes an impurity in one of the
anharmonic oscillators of the lattice\cite{s:3}. In Josephson junctions, $%
\phi $ is the phase difference of the superconducting electrons across the
junction, $F(x)$ is the external current, and $N(x)$ can describe a
microshort or a microresistor\cite{s:4}. In a Josephson transmission line it
is possible to apply nonuniformly distributed current sources ($F(x)$) and
to create inhomogeneities of type $N(x)$ using different electronic circuits
in some specific elements of the chain\cite{s:2,s:5}.

In the present paper the functions $F(x)$ and $N(x)$ will be defined as,

\begin{equation}
F(x)=\frac 12A(A^2-1)\tanh (Bx),  \label{aa:1}
\end{equation}

\begin{equation}
\ N(x)=\frac 12\frac{(4B^2-A^2)}{\cosh {}^2(Bx)}.  \label{aa:2}
\end{equation}

The case $F=const.$ has been studied in many papers (see e.g. \cite{s:1}).
Here Eq.~(\ref{aa:1}) represents an external field (or a source current in a
Josephson junction) that is almost constant in most part of the chain but
changes its sign in $x=0$ (this is very important in order to have soliton
pinning\cite{r:1}). Microshorts, microresistors or impurities in
atomic chains\cite{s:4} are usually described by Dirac's delta functions ($%
\delta (x)$) where the width of the impurity is neglected. The function $%
N(x) $ is topologically equivalent to a $\delta (x)$ but it allows us to
consider the influence of the width of the impurity.

\section{Stability analysis}

\label{sec2}

Suppose the existence of a static kink solution $\phi _k(x)$ corresponding
to a soliton placed in a stable equilibrium state created by the
inhomogeneities of Eq.~(\ref{a:1}). We analyze the small amplitude
oscillations around the kink solution $\phi (x,t)=\phi _k(x)+\psi (x,t)$. We
get for the function $\psi (x,t)$ the following equation:

\begin{equation}
\psi _{xx}-\psi _{tt}-\gamma \psi _t+\frac 12(1-3\phi _k^2+2N(x))\psi =0.
\label{a:3}
\end{equation}

The study of the stability of the equilibrium solution $\phi _k(x)$ leads to
the following eigenvalue problem (we introduce $\psi (x,t)=f(x)\exp (\lambda
t)$ into Eq.~(\ref{a:3})):

\begin{equation}
-f_{xx}+\frac 12(3\phi _k^2-1-2N(x))f=\Gamma f,  \label{a:4}
\end{equation}
where $\Gamma \equiv -\lambda ^2-\gamma \lambda $.

For the functions $F(x)$ and $N(x)$ (defined as Eqs.(\ref{aa:1}-\ref{aa:2}))
the exact solution describing the static soliton can be written: $\phi
_k(x)=A\tanh (Bx)$. The spectral problem (Eq.~(\ref{a:4})) brings the
following eigenvalues for the discrete spectrum: $\Gamma _n=\frac
12A^2-\frac 12+B^2(\Lambda +2\Lambda n-n^2-2)$; here $\Lambda $ is defined
as, $\Lambda (\Lambda +1)=\frac{A^2}{B^2}+2$. The integer part of $\Lambda $
($\left[ \Lambda \right] $) defines the number of modes of the discrete
spectrum.

The stability condition for the translational mode is, $%
16B^4+2B^2(5-7A^2)+(1-A^2)^2<0$. When this condition is not fulfilled (thus
the equilibrium point $x=0$ is unstable) and $A^2>1$, then there will exist
three equilibrium points for the soliton: two stable (at points $x=x_1>0$
and $x=x_2<0$) and one unstable at point $x=0$. This is because for huge
values of $\left| x\right| $ the leading inhomogeneity is $F(x)$, which is
non-local and not zero at infinity. This inhomogeneity acts as a restoring
force that pushes the soliton towards the point $x=0$. As a result of the
competition between the local instability induced by $N(x)\phi $ at point $%
x=0$ and the non-local inhomogeneity $F(x)$, an effective double-well
potential is created. This is equivalent to a pitchfork bifurcation.

We should make some remarks about the stability investigation. 
Writing down Eq.~(\ref{a:3}) we are making an approximation because the terms $\psi ^2$ and $\psi ^3$
are considered zero. Under this assumption the
solutions of Eq.~(\ref{a:3}) can be used as an approximation for the kink dynamics
only for small perturbation of the static soliton solution. However, the 
stability conditions obtained for the different modes are exact.
In fact, when we say that the
traslational mode is stable for some set of values of the parameters, this
means that in a neighborhood of this equilibrium point the effective
potential for the soliton center of mass is a well (a minimum). On the 
contrary, when the parameters are changed such that the stability condition does not
hold anymore, then a small deviation in the initial condition of the 
soliton center of mass will cause the soliton to move away from the
equilibrium position. The same is valid for the stability of the shape modes.
For example, if the stability condition for the first shape mode is not 
satisfied, then small perturbation of the soliton profile will cause the
soliton to explode. This has been checked numerically\cite{t:2}.

In general, the stability problems for perturbed soliton equations are
very hard\cite{s:4}. This is because in order to solve it, we first should have an
exact solution of the equilibrium problem (which is rarely the case), and 
then one should solve the eigenvalue problem which usually has no solution
in terms of elementary functions.

The investigation we have performed includes several steps. First, we have to
solve an inverse problem in order to have external perturbations with the
``shapes'' that are relevant to the physical situations we want to discuss;
second, we assure that the exact solutions will be known to us, 
and third, we should be able to solve exactly the stability problem. This last
condition is fulfilled because Eq.~(\ref{a:4}) is a Schr\"odinger equation with
a P\"oschl-Teller potential\cite{r:1,r:2,t:1}. The solution of this spectral problem can be found in Ref. \cite{t:3}.

In our case we were lucky enough to obtain exact solutions to perturbations
that are generic and topologically equivalent to well-known perturbation
models (e.g. the pitchfork bifurcation).

\section{Karhunen-Lo\`eve analysis}

Let us consider a space-time-dependent force $G(x,t)$ beside the
space-dependent forces $F(x)$ and $N(x)\phi $. In a previous work \cite{r:1}%
, Gonz\'alez and Ho\l yst found that if $G(x,t)$ has a spatial shape such
that it coincides with one of the eigenfunctions of the stability operator
of the soliton, then it is possible to get resonance if the frequency of the
force also coincides with the resonant frequency of the considered mode.
Therefore we can pump energy only into the traslational mode of the kink
selecting a space-time-dependent force of the form 
\begin{equation}
G(x,t)=\nu \cos (\omega t)\left( \frac 1{\cosh {}^\Lambda (B(x-x_1))}+\frac
1{\cosh {}^\Lambda (B(x+x_1))}\right) .  \label{b:9}
\end{equation}

In Fig. \ref{Fig.1}(a) we present a sequence of bifurcations of the soliton
center-of-mass coordinate $X_{c.m.}=\frac{\int_{-l/2}^{l/2}x\phi _x^2dx}{%
\int_{-l/2}^{l/2}\phi _x^2dx}$ (sampled at times equal to multiples of the
period of the driving force) as the driving amplitude $\nu $ is increased
and other parameters remain fixed ($A=1.22$, $B=0.32$, $\omega =1.22$, $%
x_1=2.5$ and $\gamma =0.3)$. For these values of $A$ and $B$ the stability
condition for the translational mode is fulfilled, the soliton moves in a
single-well potential and the system is in a regime with three discrete
modes ($\left[ \Lambda \right] =3$). Previous articles have studied the
bistable case as well as the single-well case created by inhomogeneities of
the type $F(x)$\cite{r:1,r:2}. In this article we want to stress the
complexity of the internal dynamics of the soliton when, besides $F(x)$,
there is an impurity of the type $N(x)\phi $.

We have integrated the equation using a standard implicit finite difference
method with open boundary conditions $\phi _x(0,t)=\phi _x(l,t)=0$ and a
system length $l=80$. We use a kink-soliton with zero velocity as initial
condition.

\begin{figure}[tbp]
\caption{(a) Bifurcation diagram for the position of the center of mass of
the soliton. (b) Relative weight of the highest KL eigenvalue. (c) Number of
KL modes that contains 99.9\% of the dynamics.}
\label{Fig.1}\vspace{100mm}
\end{figure} 

Poincar\'e maps for the soliton center-of-mass coordinate have revealed
quasiperiodic and chaotic attractors for the non-periodic solutions of Fig. 
\ref{Fig.1}(a): period one solutions precede a window of quasiperiodic
bifurcations (the torus entangles as the amplitude of the time-dependent
driving force increases). At a certain value a period two window appears and
is followed by quasiperiodic (two-tori) bifurcations. For $\nu =0.55$ the
Poincar\'e maps reveal high-dimensional chaotic motion followed by period one solutions.

The KL decomposition\cite{r:5,r:6} allows to describe the dynamics in terms
of an adequate basis of orthonormal functions or modes. The eigenvalues $%
\lambda _n$ can be regarded as the weight of the mode $n$. Figure \ref{Fig.1}%
(b) presents the the greater eigenvalue normalized by the weight, $W=\sum
\lambda _n$, whereas Fig. \ref{Fig.1}(c) presents the number of modes that
contains $99.9\%$ of the weight.

\begin{figure}[tbp]
\caption{KL spectra for the sequence of bifurcations presented in Fig. \ref
{Fig.1}. The inset shows the first mode of the KL spectrum for $\nu =0.20$
and $\nu =0.60$.}
\label{Fig.2}\vspace{60mm}
\end{figure} 

Figure \ref{Fig.2} reveals the increasing excitation of the KL modes as the
amplitude of the space-time-dependent force increases. Note the sudden
changes of the spectra when periodic motion is regained (period-two for $\nu
=0.40$ and period-one for $\nu =0.60$). For these solutions the amplitude of
the oscillations around the point $x=0$ diminishes even though the amplitude
of the driving force has increased. This agrees with the higher contribution
to the dynamics of the few modes of shape whereas all the rest of the modes
decreased their contribution. Furthermore, for $\nu =0.60$ the first shape
mode replaces the translational mode as the leading mode of the dynamics.
The inset of the Figure \ref{Fig.2} presents the leading KL eigenmodes for
the period-one solutions that initiates and ends the sequence of
bifurcations considered in this section. The eigenmode for $\nu =0.20$
appears to be the superposition of a pair of translational modes centered at
the equilibrium points for the soliton. Similar situation occurs for $\nu
=0.60$ but the eigenvalue appears to be the superposition of a pair of shape
modes.

This work has been partially supported by Consejo Nacional de
Investigaciones Cient\'{i}ficas y Tecnol\'ogicas (CONICIT) under Project
S1-2708.

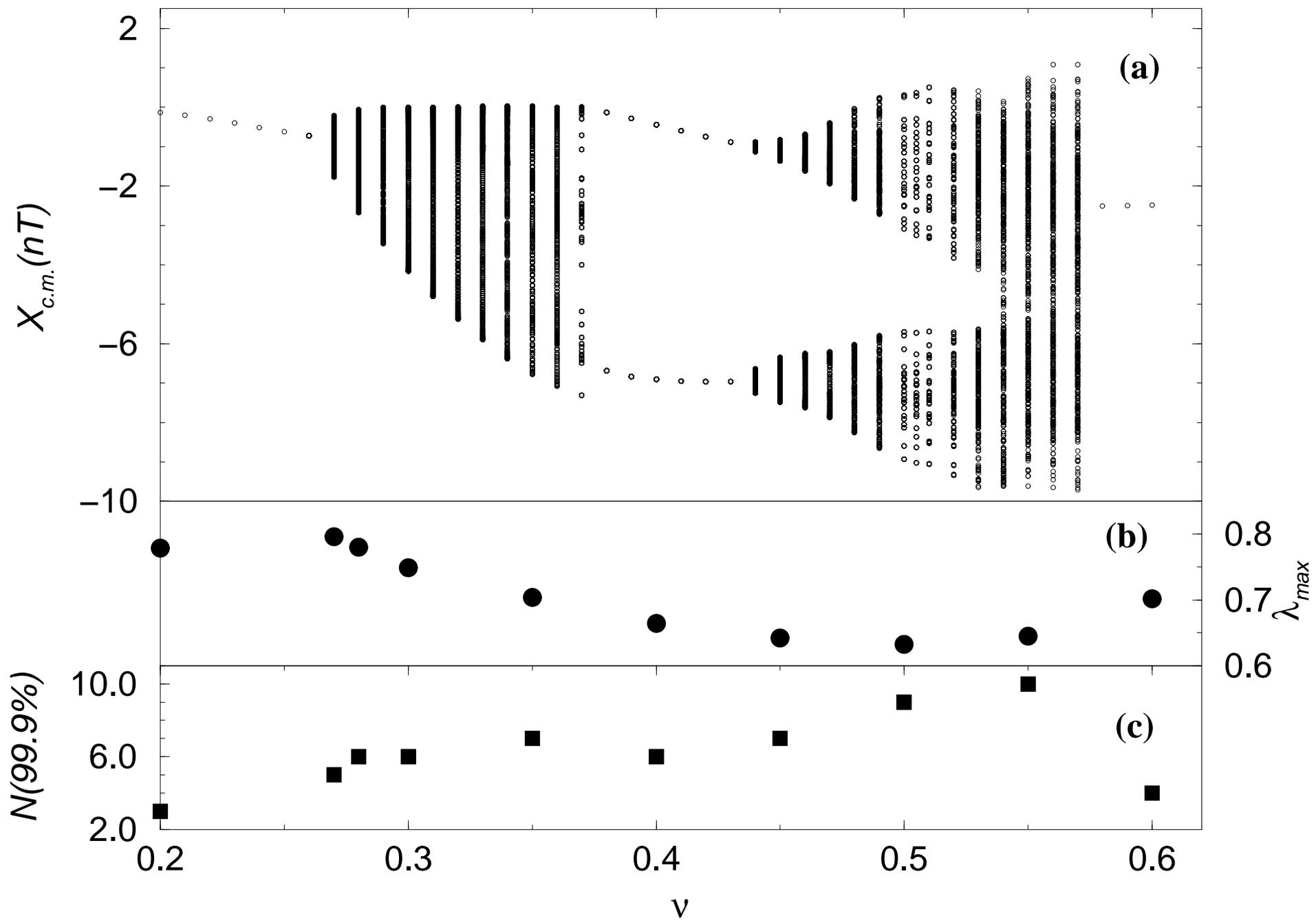
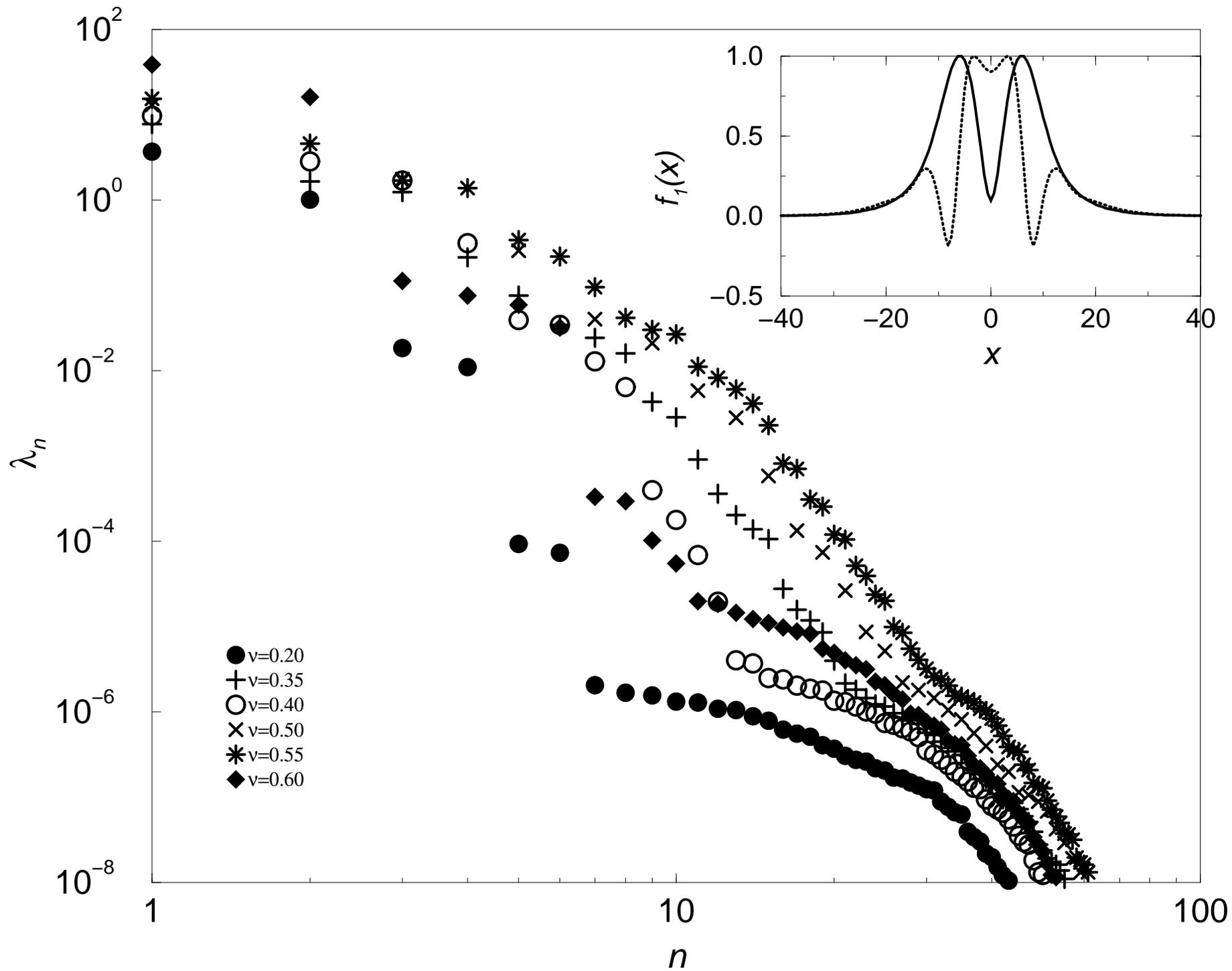
\end{document}